\documentclass[runningheads,orivec]{llncs}
\usepackage[T1]{fontenc}
\usepackage{graphicx}
\usepackage{caption}
\usepackage{amsmath,amssymb,amsfonts}
\usepackage{mathtools}
\usepackage{thmtools}
\usepackage{bm}
\usepackage{adjustbox}
\usepackage{listings}
\usepackage{color, xcolor}
\usepackage{algorithmic}
\usepackage[ruled,vlined,linesnumbered]{algorithm2e}
\usepackage{graphicx}
\usepackage{textcomp}
\usepackage{url}
\usepackage{xspace}
\usepackage{multirow}
\usepackage{subcaption} 
\usepackage{wrapfig} 
\usepackage{framed}
\usepackage{enumitem}
\usepackage{hyperref}
\usepackage{marvosym}

\definecolor{codegreen}{rgb}{0,0.6,0}
\definecolor{codegray}{rgb}{0.5,0.5,0.5}
\definecolor{codepurple}{rgb}{0.58,0,0.82}
\definecolor{backcolour}{rgb}{0.97,0.97,0.97}
\definecolor{lightgray}{rgb}{.9,.9,.9}
\definecolor{darkgray}{rgb}{.4,.4,.4}
\definecolor{promptblue}{rgb}{0.0,0.2,0.6}
\definecolor{promptgreen}{rgb}{0.0,0.45,0.0}
\definecolor{promptorange}{rgb}{0.75,0.35,0.0}

\lstdefinelanguage{SMT}{
keywords={ 
  declare, datatypes, declare, fun, property
  },
ndkeywords=[2]{assert, forall, set, logic, check, sat},
morekeywords=[3]{plus, mult, leq},
escapeinside={(*}{*)}, 
keywordstyle=\color{blue},
keywordstyle=[2]\color{codepurple},
keywordstyle=[3]\color{orange!80!black},
identifierstyle=\color{black},
sensitive=true,
comment=[l]{;},
commentstyle=\color{codegreen}\ttfamily,
stringstyle=\color{red}\ttfamily
}

\lstdefinelanguage{Prompts}{
  keywords={Task,Environment,Description,Requirements,Input,format,Chain,Thought,Output,file,SMTLIB2},
  sensitive=false,
  keywordstyle=\color{promptblue}\bfseries,
  identifierstyle=\color{black},
  comment=[l]{\#},
  commentstyle=\color{codegreen}\ttfamily,
  stringstyle=\color{red}\ttfamily
}
\lstdefinestyle{Prompts}{
  language=Prompts,
  backgroundcolor=\color{backcolour},
  extendedchars=true,
  basicstyle=\footnotesize\ttfamily,
  showspaces=false,
  numbers=left,
  numberstyle=\tiny\color{codegray},
  numbersep=10pt,
  tabsize=2,
  breaklines=true,
  showtabs=false,
  captionpos=b,
  breakatwhitespace=false,
  keepspaces=true
}

\lstdefinestyle{SMT}{
language=SMT,
backgroundcolor=\color{white},
extendedchars=true,
basicstyle=\footnotesize\ttfamily,
keywordstyle=\color{blue},
showspaces=false,
numbers=left,
numberstyle=\tiny\color{codegray},
numbersep=10pt,
tabsize=2,
breaklines=true,
showtabs=false,
captionpos=b,
breakatwhitespace=false,
keepspaces=true
}

\lstdefinestyle{inlineSMT}{
language=SMT,
backgroundcolor=\color{white},
extendedchars=true,
basicstyle=\footnotesize\ttfamily,
keywordstyle=\color{blue},
showstringspaces=false,
showspaces=false,
numbers=left,
numberstyle=\tiny\color{codegray},
numbersep=2pt,
tabsize=2,
breaklines=true,
showtabs=false,
captionpos=b,
breakatwhitespace=true,
keepspaces=true
}

\newcommand{\tool}{{\sc LLM4Ind}\xspace}

\newcommand{\code}[1]{{\lstinline[language=SMT,style=inlineSMT]{#1}}\xspace}

\newcommand{\hide}[1]{}
\newif\iffullversion
\fullversiontrue   
\newcommand{\toggle}[2]{%
  \iffullversion
    #1%
  \else
    #2%
  \fi
}

\begin{document}
\title{Can LLM Aid in Solving Constraints with Inductive Definitions?}
%
%
\author{Weizhi Feng\inst{1,2}\orcidID{0000-0003-0710-223X} 
\and
Shidong Shen\inst{1,2}\orcidID{0009-0000-0369-021X} 
\and
Jiaxiang Liu\inst{1,2}\orcidID{0000-0002-6725-8167}
\and
Taolue Chen\inst{3}\orcidID{0000-0002-5993-1665}
\and\\
Fu Song\inst{1,2,4}\orcidID{0000-0002-0581-2679}
\and
Zhilin Wu\inst{1,2}\textsuperscript{\raisebox{0.2ex}{(\Letter)}}\orcidID{0000-0003-0899-628X}
}
\authorrunning{W. Feng et al.}
%
\institute{Key Laboratory of System Software (Chinese Academy of Sciences),
Institute of Software, Chinese Academy of Sciences, Beijing, China 
\and
University of Chinese Academy of Sciences, Beijing, China
\and
Birkbeck, University of London, United Kingdom 
\and
Nanjing Institute of Software Technology, Nanjing, China \\
\email{\{fengwz,shensd,liujx,songfu,wuzl\}@ios.ac.cn \quad 
t.chen@bbk.ac.uk}
}
\maketitle              
\begin{abstract}
Solving constraints involving inductive (aka recursive) definitions is challenging. 
State-of-the-art SMT/CHC solvers and first-order logic provers provide only limited support for solving such constraints, especially when they involve, e.g., abstract data types. 
In this work, we leverage structured prompts to elicit Large Language Models (LLMs) to generate auxiliary lemmas that are necessary for reasoning about these inductive definitions. We further propose a neuro-symbolic approach, which synergistically integrates LLMs with constraint solvers: the LLM iteratively generates conjectures, while the solver checks their validity and usefulness for proving the goal. We evaluate our approach on a diverse benchmark suite comprising constraints originating from algebraic data types and recurrence relations. The experimental results show that our approach can improve the state-of-the-art SMT and CHC solvers, solving considerably more (around 25\%) proof tasks involving inductive definitions, demonstrating its efficacy.

\end{abstract}

\section{Introduction}

Inductive definitions are prevalent in program verification. Typically, they exhibit in two forms, i.e.,  algebraic data types (ADTs) and recursively defined functions (RDFs). An ADT defines a type whose elements are constructed by a finite number of applications of a given set of rules.
\begin{lstlisting}[language=SMT, style=SMT,
  label=lst:adt-nat-example, float=tp, belowcaptionskip=-20pt,
  caption={A pseudocode program: the running example}]
datatypes Nat := zero | succ(n: Nat) (*\label{lst:adt-nat}*)
fun plus(Nat, Nat): Nat { (*\label{lst:fun-plus-begin}*)
  (*$\forall$*)y(*$\in$*)Nat.plus(zero,y) = y
  (*$\forall$*)x(*$\in$*)Nat,y(*$\in$*)Nat.plus(succ(x),y) = succ(plus(x,y))
} (*\label{lst:fun-plus-end}*)
fun mult(Nat, Nat): Nat { (*\label{lst:fun-mult-begin}*)   
  (*$\forall$*)y(*$\in$*)Nat.mult(zero,y) = zero
  (*$\forall$*)x(*$\in$*)Nat,y(*$\in$*)Nat.mult(succ(x),y) = plus(mult(x,y),y)
} (*\label{lst:fun-mult-end}*)
property: (*$\forall$*)x(*$\in$*)Nat,y(*$\in$*)Nat.mult(x,y) = mult(y,x) (*\label{lst:property}*)
\end{lstlisting}
%
As an example, Listing~\ref{lst:adt-nat-example} 
defines natural numbers (\code{Nat}) as an ADT, and two operations on natural numbers as RDFs. 
\code{Nat} is defined as either zero (\code{zero}) or the successor of another \code{Nat}, given by \code{succ(n)}. 
Both RDFs \code{plus} and \code{mult} are
specified through assertions with universal quantifiers,  serving as axioms. 
\begin{itemize}[topsep=0.2em,leftmargin=*]
    \item The addition operation (\code{plus}) (lines~\ref{lst:fun-plus-begin}--\ref{lst:fun-plus-end}) is recursively defined by two cases: (i)~adding \code{zero} to any natural number \code{y} yields \code{y}, and (ii)~adding the successor of any natural number \code{x} to \code{y} yields the successor of \code{plus(x,y)}.
    \item The multiplication operation (\code{mult}) (lines~\ref{lst:fun-mult-begin}--\ref{lst:fun-mult-end}) is recursively defined by two cases as well:
    (i)~multiplying \code{zero} by any natural number \code{y} results in \code{zero}, and 
    (ii)~multiplying the successor of any natural number \code{x} by \code{y}  is the addition of \code{mult(x,y)} and \code{y}.
\end{itemize}
In this example, the proof goal is the commutativity of the multiplication operation \code{mult}, formulated as 
\code{property} at line~\ref{lst:property} in Listing~\ref{lst:adt-nat-example}.

Such ADTs are ubiquitous in 
functional programming languages (e.g., Gallina) for  
interactive  theorem provers (ITPs, e.g., Rocq~\cite{bertot2013interactive}), 
and are increasingly being supported by other languages such as TypeScript and Rust.

\smallskip
\noindent{\textbf{Inductive Reasoning.}} 
Program verification requires reasoning about properties 
of inductive definitions, including ADTs and RDFs. 
The standard reasoning process typically consists of two steps: first proving the base case(s) and then proving the inductive case(s) with inductive hypotheses as premises. For example, in Listing~\ref{lst:adt-nat-example}, one may apply induction on the natural number \code{x}, leading to the base case where \code{x} is \code{zero} and an inductive case where
\code{x} is \code{succ(x')} in which the property is assumed to hold for \code{x'}. 
These two cases yield two verification conditions (VCs):
\begin{equation} 
\label{eq:twocases} 
\begin{aligned}
\forall \ y \in \texttt{Nat}. \  \texttt{mult}(\texttt{zero}, y) & = \texttt{mult}(y, \texttt{zero}), \\
\forall \ x', y \in \texttt{Nat}. \ \texttt{mult}(x', y) & = \texttt{mult}(y, x') \rightarrow \\ 
&\texttt{mult}(\texttt{succ}(x'), y) = \texttt{mult}(y, \texttt{succ}(x')).
\end{aligned}
\end{equation}
Different techniques can be applied for inductive reasoning, 
ranging from manual to fully automated. In general, ITPs such as Rocq~\cite{bertot2013interactive}, Isabelle~\cite{paulson1994isabelle} and Lean~\cite{Moura021}, 
and semi-automated verifiers (aka deductive verifiers) such as Dafny~\cite{Leino10}, Stainless~\cite{HamzaVK19} and Verus~\cite{LattuadaHCBSZHPH23}, require 
user intervention. 

In contrast, automated constraint solving, including SMT solving and first-order logic theorem proving,
is designed to operate 
without user intervention. Typically, induction schemas are incorporated
into SMT solvers~\cite{ReynoldsK15} or into superposition-based first-order logic theorem provers~\cite{Cruanes17,RegerV19,DBLP:journals/jar/EchenimP20,HajduHKSV20,HajduHKV21,HozzovaKV21,HajduKRV22}, enabling automatic generation of base and inductive cases as verification conditions with inductive hypotheses as premises. However, many proof goals with inductive definitions cannot be solved by solely applying the inductive hypotheses and axioms. Consequently, additional \emph{auxiliary lemmas} are required to complete the proof. Namely, to prove $A \rightarrow P$ where $A$ and $P$ are the premise and the conclusion, respectively, we may introduce a set of auxiliary lemmas $\{L_i\}_{i\in I}$ such that 
$A \rightarrow L_i$ holds for each $i\in I$, and  
$A \wedge \bigwedge_{i\in I} L_i \rightarrow P$ holds. 

\smallskip 
\noindent{\textbf{Lemma Generation Methods and Limitations.}} This paper tackles \emph{automated generation of auxiliary lemmas} in constraint solving. The existing methods can roughly be classified into three main categories. 
First, \emph{theory exploration} method, implemented in solvers such as cvc5~\cite{BarbosaBBKLMMMN22}, iteratively enumerates conjectures starting from small terms and applies heuristic checks to quickly filter out invalid ones before verifying their validity and usefulness for assisting in proving the goal~\cite{ClaessenJRS13,ReynoldsK15,SingherI20}.
Second, the \emph{generalization} method, implemented in solvers such as Vampire~\cite{KovacsV13},  identifies common sub-terms in the proof goal and replaces them with new variables to find auxiliary lemmas that are simpler and can generalize the goal~\cite{HajduHKSV20}.
Third, \emph{Constrained Horn Clauses (CHCs) based} method~\cite{GovindShohamGurfinkel2022} treats inductive definitions as transition systems and solves the verification problem by synthesizing inductive invariants. (Cf.\ Section~\ref{sect:related} for a brief review of the related work.) 
 
Despite their effectiveness in certain scenarios, these methods have significant limitations. Theory exploration is effective for generating simple lemmas but often struggles with discovering more complicated lemmas required for inductive proofs. Indeed, the representative state-of-the-art SMT solver, cvc5, fails to prove the 
\code{property} given in Listing~\ref{lst:adt-nat-example} 
\toggle{(cf.\ Appendix~\ref{sec:examplecvc5} for details).}{(cf.\ the full version of the paper for details).}
Generalization methods have limited expressiveness and cannot handle more general problems. CHC-based methods have limited capability to handle inductive definitions, particularly RDFs. 
Overall, these traditional logic-based methods mostly rely on fixed mechanisms and heuristic strategies, which often suffer from limited scalability or expressiveness, restricting the usability and applicability 
of automated constraint solving in program verification.

\smallskip 
\noindent{\textbf{LLM-aided Solving.}}
LLMs 
have 
demonstrated remarkable capabilities in code generation~\cite{guo2024deepseek,sun2025don,sun2024neural,Yang2024ChainOfThought}, program specification inference~\cite{wen2024enchanting,specGen},   
loop-invariant generation~\cite{CaoWXYWCM25,WuC0W0M24,PirzadaRBC24}, 
and interactive theorem proving~\cite{deepseek-prover-v2,kimina-prover}. 
Recent works have also explored LLM-assisted lemma/conjecture generation in deductive verification (e.g., Dafny~\cite{SilvaMF24} and Verus~\cite{YangLMYCGHLLL0Z25}), hardware model checking~\cite{PeledKTV25} and mathematical reasoning~\cite{AlhessiEGFLJS25,VaramballyVSCYY25,ChuharskiCM24}.
Overall, these methods largely follow a generate-then-verify paradigm,
but they target either program-level proof construction or hardware/mathematical formula reasoning, rather than fully automated inductive reasoning and first-order logic constraint solving. 

Inspired by these advances of LLMs in generative AI and formal reasoning, we introduce LLMs into automated inductive reasoning, in particular, to overcome the limitations of pure logic-based methods in lemma generation.
However, there are two main technical challenges:

\begin{description}[topsep=0.2em]
    \item {\bf Challenge~1.} While LLMs often exhibit emergent abilities across diverse problem types, they may not be very effective in specific downstream tasks. The primary challenge is to effectively unleash their capability for inductive reasoning, e.g., to guide them to recognize what constitutes \textit{good} auxiliary lemmas therein. 
    \item {\bf Challenge~2.} Unlike traditional logic-based methods, the outputs of LLMs exhibit randomness, and can even hallucinate. 
The second challenge is how to 
quickly eliminate invalid outputs and 
ensure their validity and 
usefulness for assisting in proving the target goal. 
\end{description}
  
\noindent{\textbf{Our Contributions.}} 
We propose a neuro-symbolic approach that synergistically integrates LLMs and constraint solving for automated lemma generation in inductive reasoning. Our approach consists of three stages: \textit{query}, \textit{filter} and \textit{validate}. 
Given an instance of the inductive reasoning problem that comprises inductive definitions and a proof goal, our approach automatically generates auxiliary lemmas and proves 
the goal. The three stages are designed to  
address the above two technical challenges. To address Challenge 1,
we design two prompt strategies in the query stage:
(i) the first strategy imitates human inductive reasoning through equational term rewriting, enabling goal-oriented lemma discovery, and 
(ii) the second strategy abstracts the proof goal into simpler forms and generates bridging lemmas between simplified forms and the original goal. Moreover, it incorporates heuristics, such as suggesting commonly used axioms, attempting to generate strengthened propositions that imply the original goal, and applying equivalent transformations to terms in the proof goal.
The filter and validate stages together address Challenge 2, where the filter stage quickly eliminates incorrect and useless outputs using backend solvers with short timeouts, resulting in candidate lemmas, while the validate stage enforces validity and usefulness of candidate lemmas through a two-step process: checking whether the candidate lemmas can assist in proving the original goal, and then recursively verifying each candidate lemma if they suffice to prove the goal.

We evaluate our approach on 706 instances of the inductive reasoning problem collected from diverse benchmarks commonly used in the inductive reasoning literature~\cite{ReynoldsK15,GovindShohamGurfinkel2022,SunJFJCX24}. 
We compare our method against state-of-the-art solvers (including cvc5 and Vampire). 
The experimental results show that our approach achieves (at least 25\%) higher success rate than these solvers. 
Additionally, the ablation studies show that both the filtering and prompt designs contribute substantially to the performance improvement. Further experiments also suggest that our approach is robust across different LLMs, sampling temperatures and backend solvers, 
which demonstrates the generality of our approach.

\toggle{The tool, benchmarks and experimental data in this paper is available at \url{https://github.com/fengwz17/LLM4Ind}.}
{The full version of the paper, as well as the tool, benchmarks and experimental data, is available at \url{https://github.com/fengwz17/LLM4Ind}.}
\section{Preliminary}
\label{sec:pre}
\subsection{Satisfiability Modulo Theories}

Satisfiability Modulo Theory (SMT) refers to deciding whether a given logical formula is satisfiable, i.e., whether there exists an assignment of its variables over background theories, under which the formula holds.  

Formally, an SMT formula is defined over a given signature
$\Sigma$ that consists of a set of sorts (types) $\mathcal{S}$, function symbols $\mathcal{F}$ and predicate symbols $\mathcal{P}$. 
A model $\mathcal{M}$ of a given theory $\mathcal{T}$ (e.g., Linear integer arithmetic (LIA), Bit-vector (BV), Uninterpreted functions (UFs), Algebraic data types (ADTs)) provides an interpretation of sorts, function symbols and predicate symbols in $\Sigma$.
\hide{
A formula $\varphi$ holds (or is satisfied, or is true) in a model $\mathcal{M}$, written $\mathcal{M} \models_{\mathcal{T}} \varphi$, if the interpretation of $\varphi$ under $\mathcal{M}$ evaluates to true. 
A formula $\varphi$ is satisfiable in $\mathcal{T}$ if there exists some $\mathcal{M} \models_{\mathcal{T}} \varphi$; otherwise, it is unsatisfiable. 
A formula $\varphi$ is said to be valid in $\mathcal{T}$, denoted $\models_{\mathcal{T}} \varphi$, if $\mathcal{M} \models_{\mathcal{T}} \varphi$ for every model $\mathcal{M}$ of $\mathcal{T}$; in this case, its negation $\neg \varphi$ is unsatisfiable. If a formula $\varphi$ does not hold in a given model $\mathcal{M}$, we write $\mathcal{M} \not\models_{\mathcal{T}} \varphi$; if there exists at least one such model, we say $\varphi$ is invalid in $\mathcal{T}$.
}
A  set of clauses $F = \{ C_1, \cdots, C_n \}$ represents the conjunction 
$C_1 \wedge \cdots \wedge C_n$, where each clause $C_i$ may be universally quantified. 

SMT solvers, such as Z3~\cite{MouraB08}, cvc5~\cite{BarbosaBBKLMMMN22} or Yices~\cite{Dutertre14}, combine SAT-solving techniques with specialized theory solvers. The SAT engine handles the propositional structure of the formula, while theory solvers reason about constraints in their respective domains. Through this integration, SMT solvers have become powerful tools for, among others, program verification and synthesis, model checking, and test-case generation. 
Besides, some first-order logic provers, such as Vampire~\cite{KovacsV13}, can also accept a fragment of SMTLIB2 inputs~\cite{RegerSV17}. For simplicity, we refer to both SMT solvers and such first-order logic provers as \emph{solvers} in this paper when no ambiguity arises.

\subsection{Inductive Definitions}
{\bf Algebraic Data Types.} 
The theory of algebraic data types (ADTs) is defined over a signature $\Sigma$.
For example, natural numbers and lists can be defined as:
\begin{center}
    \begin{tabular}{l}
     \code{datatypes Nat := zero | succ(n: Nat)},   \\
     \code{datatypes List := nil | cons(head: Nat, tail: List)}. 
\end{tabular}
\end{center}
$\Sigma$ includes sorts for representing ADTs (e.g., \code{Nat}, \code{List}), 
function symbols for constructors (e.g., \code{zero}, \code{succ} for \code{Nat}) and selectors (e.g., \code{head}, \code{tail} for \code{List}), and predicate symbols for testers.

\smallskip\noindent
{\bf Recursively Defined Functions.}
A recursively defined function (RDF) is a function whose definition refers to itself on smaller or simpler inputs. Formally, such a function is specified by a set of base cases, which provide values for the simplest inputs, together with recursive rules that reduce larger or more complex inputs to instances of the same function on smaller arguments. This structure ensures that every call to the function eventually reaches a base case, guaranteeing well-definedness and termination. RDFs are fundamental in mathematics and computer science, as they naturally capture inductive structures such as sequences, trees and lists. 

\subsection{Lemma Generation}
\label{ssec:lemmaGeneration} 
Let $A$ denote a set of axioms encoding RDFs 
over ADTs, and $P$ denote a target property to be proved.
$A$ usually consists of universally quantified equations, 
while $P$ is a universally quantified formula, 
e.g., $\forall \vec{x}\in\tau.\ t(\vec{x})$, where $\vec{x} = \{ x_1, \cdots, x_n \}$, $\tau$ is an ADT sort (e.g., \code{Nat}), and $t$ is a term over $\vec{x}$.
The goal is to establish the validity of $A \rightarrow P$. To this end, we aim to generate a set of lemmas $\mathcal{L} = \{ L_i \}_{i \in I}$ such that the following conditions hold:
\[
    A \wedge \bigwedge_{i\in I} L_i \rightarrow P, \quad \text{and} 
    \quad \forall i \in I. \ \ A \rightarrow L_i. 
\]
%
In practice, to prove $A \rightarrow P$, SMT-based approaches negate the property $P$, and check whether the formula, given by 
the set of clauses $F = \{A,  \neg P\}$, is unsatisfiable.
Then the verification task is checking the satisfiability of $F$. 
The \textit{lemma generation task} is to find a set of lemmas $\mathcal{L} = \{ L_i\}_{i \in I}$ such that:
(i) each lemma $L_i$ is entailed by $A$, i.e., $\{ A, \neg L_i \}$ is unsatisfiable, and
(ii) the set of clauses $\{ A, \neg P \} \cup \mathcal{L}$ is unsatisfiable.

SMT solving commonly copes with universally quantified formulas such as $\forall \vec{x} \in \tau. t(\vec{x})$ through instantiation-based techniques, and handles the negation of universally quantified formulas (effectively existentially quantified formulas since $\neg \forall \vec{x} \in \tau. t(\vec{x})$ is equivalent to $\exists \vec{x} \in \tau. \neg t(\vec{x})$ ) via skolemization. 
cvc5 implements an inductive strengthening technique in its skolemization module to incorporate the induction schema into SMT solving
and a theory exploration method to automatically
generate auxiliary lemmas.
There are other solvers
such as Vampire~\cite{HajduKRV22} and Racer~\cite{GovindShohamGurfinkel2022} 
which support 
reasoning about inductive definitions
and are reported to outperform cvc5 in some categories of benchmarks. However, in general,
cvc5 represents the state-of-the-art 
in 
inductive reasoning, which is used as the backend solver. 

Hereafter, we adopt the following terminology for the LLM-aided lemma generation task. 
Outputs generated by LLMs are referred to as \emph{conjectures} if they are in the correct 
format. 
Conjectures $\{ L_i\}_{i \in I}$  are said \emph{useful} 
if they can assist in proving $P$ when used as premises, namely, $A \wedge \bigwedge_{i\in I} L_i \rightarrow P$ can be proved. 
A conjecture $L$ is referred to as a \emph{lemma} if it can be proved under the axiom $A$. 
%
Useful conjectures $\{ L_i\}_{i \in I}$ are referred to as 
\emph{auxiliary lemmas} when all of them can be proved under the axiom $A$, i.e., $A \rightarrow L_i$ can be proved for all $i\in I$. To establish the validity of $A \rightarrow P$, our main goal is to elicit LLMs to generate auxiliary lemmas.

\section{Our Neuro-Symbolic Approach}
\label{sec:approach}

\subsection{Motivating Example}
\label{subsec:motivating-example}

We first illustrate the challenges of naively querying LLMs for lemma generation without carefully designed prompts. 
Consider the task of verifying the property in Listing~\ref{lst:adt-nat-example} using the following naive prompt:

\begin{lstlisting}[language=SMT, style=SMT, label=lst:naive-prompts, caption={A naive prompt for lemma generation}]
You are an expert in constraint solving, inductive reasoning, and functional program verification. 
Please generate auxiliary lemmas in SMTLIB2 format to add to the following file, which can help the solver verify the property.

  { Input SMTLIB2 file }
\end{lstlisting}

Due to randomness and possible hallucination, LLM-generated conjectures may not be auxiliary lemmas. Based on our experiment, these conjectures can be categorized into the following three types:

\begin{itemize} [topsep=0.2em,leftmargin=*]
    \item {\bf Incorrect Conjectures.} 
    When the outputs of LLMs are complete and syntactically correct,
    forming conjectures, 
    they may be semantically incorrect. Namely,
    an LLM-generated conjecture $L$ contradicts the axiom $A$, making the premise $A \wedge L$ false. In this case, 
    while $A \wedge L \rightarrow P$ is vacuously true, the conjecture $L$ does not hold under the axiom $A$ and is useless. For instance, LLMs may generate the following conjecture for the motivating example:
\[
\forall \ x \in \texttt{Nat}. \  \texttt{plus}(x, \texttt{zero}) = \texttt{zero} 
\]
which contradicts the axiom that $\texttt{plus}(x, \texttt{zero}) = x$.
   \item  {\bf Correct but Useless Conjectures.} 
   Even when an LLM-generated conjecture $L$ is correct under the axiom $A$, i.e., $A\rightarrow L$ holds, it may still be useless for proving the property $P$. For instance, an LLM generates a conjecture $L$ that is identical to the property $P$   
   or a generic property that is completely irrelevant to the property $P$.  For instance, LLMs may generate
   the following conjectures for the motivating example:
\begin{equation*} 
\begin{aligned}
\label{eq:llm-native-prompts-fail-lemmas}
&\forall \ x, y, z \in \texttt{Nat}. \ 
\texttt{plus}(\texttt{plus}(x, y), z) = \texttt{plus}(x, \texttt{plus}(y, z)), \\
&\forall \ x, y \in \texttt{Nat}. \  \texttt{plus}(x, y) = \texttt{plus}(y, x), \qquad
\forall \ x \in \texttt{Nat}. \  \texttt{mult}(x, \texttt{zero}) = \texttt{zero}).
\end{aligned}
\end{equation*}
These conjectures, while valid,  provide no assistance in proving $P$.  

\item  {\bf Useful but Remaining to be Proved Conjectures.} 
While LLMs can generate useful conjectures for proving the property $P$, these conjectures themselves should be proved from the axiom $A$, to qualify as auxiliary lemmas. In practice, 
proving useful conjectures themselves is non-trivial, e.g., some cannot be directly proved via SMT solving, leaving them as new proof obligations. 
For instance,  for the motivating example, LLMs may generate the following useful conjecture:
\[
\forall \ x, y \in \texttt{Nat}. \ \texttt{mult}(x, \texttt{succ}(y)) = \texttt{plus}(\texttt{mult}(x, y), x).
\]
This conjecture cannot be directly proved via SMT solving using cvc5, and thus requires further verification with additional auxiliary lemmas.
\end{itemize}

\smallskip
\noindent

This example demonstrates that state-of-the-art LLMs have the potential to understand inductive reasoning and generate useful conjectures beyond existing fixed mechanisms and heuristic strategies. (Note that this example cannot be proved by the representative state-of-the-art solver cvc5; cf.\   \toggle{Appendix~\ref{sec:examplecvc5}}
{the full version of the paper}.)
However, LLMs still lack clear reasoning strategies and may generate conjectures which are incorrect, useless, or themselves non-trivial to prove. 
Effectively leveraging LLMs requires carefully designed prompts to guide lemma generation and a systematic workflow to validate and integrate these conjectures, which motivates our approach. 

\subsection{Workflow}
To bring the best of LLM-aided solving and traditional logic-based solving, we propose a neuro-symbolic approach, which synergistically integrates
LLMs and SMT solvers to iteratively generate conjectures, check their
validity and usefulness for assisting in proving the target property. 
After preprocessing the input file in SMTLIB2 format, we invoke the \texttt{ProveRun} function as the main workflow to prove
the property.

\smallskip
\noindent
{\bf Preprocessing.}
Given an instance of the inductive reasoning problem as
an SMTLIB2 file, we parse the input file and organize it as three parts: datatype definitions, recursive function definitions, and the verification target, each of which is labeled with a comment (e.g., \code{; datatype definitions}, \code{; function definitions}, and
\code{;  proof goal}) to help LLMs' understanding and reasoning.

\begin{figure}[t]
    \centering
\begin{algorithm}[H]
\small
\caption{Main Workflow}
\label{alg:overall-algorithm}
\SetKwProg{Fn}{Function}{:}{}
\SetKwFunction{ProveRun}{ProveRun}
\SetKwFunction{Preprocess}{Preprocess}
\SetKwFunction{Query}{Query}
\SetKwFunction{IsFiltered}{isFiltered}
\SetKwFunction{Prove}{Prove}
\SetKwFunction{InitialCheck}{initialCheck}
$max\_depth \gets 3$, $max\_iter\_number \gets 3$; \tcp{In default} 
\tcp{Input: labeled SMTLIB2 file $P$ and current recursion depth $d$;}
\tcp{Output: {\tt True} if proved, {\tt False} otherwise;}
\Fn{\ProveRun{$P$, $d$}}{ \label{alg-line:proveRun-function}
    \If{\InitialCheck{$P$}}{ \label{alg-line:initialcheck}
        \Return {\tt True};
    }
    \lIf{$d \geq max\_depth$}{ 
        \Return {\tt False} 
    }
    \ForEach{${\tt prompt} \in {\tt prompt\_pool}$}{ \label{alg-line:prompt-strateges-begin}
        \For{$i \gets 1$ \KwTo $max\_iter\_number$}{ \label{alg-line:max-iter-numbers}
            $r, {\tt subgoals} \gets$ \Prove{$P$, {\tt prompt}}\; \label{alg-line:prove-function}
            \If{$r = {\tt True}$}{ \label{alg-line:retIsTure}
                    $all\_success \gets {\tt True}$\;
                    \textbf{parallel for} ${\tt subgoal} \in {\tt subgoals}$ \textbf{do} { \\ \label{alg-line:parallel-begin}
                    \Indp 
                        \If{$\neg$\ProveRun{{\tt subgoal}, $d+1$}}{
                            $all\_success \gets  {\tt False}$\;
                            \textbf{break}; 
                        }
                    \Indm
                    } \label{alg-line:parallel-end}
                    \If{$all\_success$}{
                        \Return {\tt True}; \label{alg-line:proveParallelReturnTrue}
                    }
            }
        }
    } \label{alg-line:prompt-strateges-end}
    \Return {\tt False}; \label{alg-line:proveRunIsFalse}
}
\end{algorithm} 
\vspace{-5mm}
\end{figure}

\smallskip
\noindent{\textbf{The \texttt{ProveRun} function.}} 
The main workflow is implemented in the \texttt{ProveRun} function in Algorithm~\ref{alg:overall-algorithm}. It takes a labeled SMTLIB2 file $P$ and the current recursion depth $d$ (initialized to $0$) as input, attempts to prove $P$ (for clarity, the property specified in $P$ is also referred to as $P$).
The recursive execution of \texttt{ProveRun} forms a proof tree
whose depth and width are bounded by the configurable parameters $max\_depth$ and $max\_iter\_number\times |{\tt prompt\_pool}|$,
where $max\_depth$ bounds the depth of recursive calls,
$max\_iter\_number$ bounds the number of queries to LLMs for each prompt, and {\tt prompt\_pool} comprises the designated prompts.

In detail, \texttt{ProveRun} first checks whether $P$ can be \emph{directly} proved via SMT solving, without LLM-generated conjectures. If it cannot be proved and the recursion has not yet reached the depth bound $max\_depth$, 
\texttt{ProveRun} attempts to prove $P$ using
designated prompts from the prompt pool
through three key stages \emph{query}, \emph{filter} and \emph{validate}.

Each prompt is queried up to $max\_iter\_number$
times (three times in this work, considering the cost and effectiveness),
at lines~\ref{alg-line:prompt-strateges-begin}--\ref{alg-line:prompt-strateges-end}. In this work, we design two effective prompt strategies (cf. Section~\ref{ssec:prompt}) to form the prompt pool, 
while other new prompt strategies can be easily extended into our workflow.

For each query of a prompt {\tt prompt}, \texttt{ProveRun} invokes
the function \texttt{Prove} (line~\ref{alg-line:prove-function}), which attempts to prove $P$ assisted by conjectures generated by LLMs using {\tt prompt}.
If $P$ is proved by \texttt{Prove} using {\tt prompt}, with LLM-generated conjectures $\{L_i\}_{i\in I}$, 
i.e., $A \wedge \bigwedge_{i\in I} L_i \rightarrow P$ is proved,
then these conjectures $\{L_i\}_{i\in I}$ are regarded as sub-goals and are validated by recursively invoking \texttt{ProveRun} in parallel. We can complete the proof of $P$ if all the sub-goals
can be proved, i.e., $A \rightarrow  L_i$ for $i\in I$ are proved.

\smallskip
\noindent\textbf{The \texttt{Prove} function}. 
The \texttt{Prove} function (Algorithm~\ref{alg:prove-algorithm}) first queries the specified LLM using ${\tt prompt}$, which will produce a set of conjectures $C$.
The LLM-generated conjectures are iteratively checked by invoking \texttt{isFiltered} (line~\ref{alg-line:filtering}), which is designed to quickly identify incorrect and useless conjectures.
If none of these conjectures can be filtered out, 
\texttt{Prove} invokes the \texttt{Verify} function (line~\ref{alg-line:verifyWithCandidateLemmas})
which checks whether these conjectures suffice to prove $P$ (line~\ref{alg-line:verifyWithCandidateLemmas}), i.e., checking the unsatisfiability of $\{A, \neg P\} \cup C$.
If this is the case  (line~\ref{alg-line:verifyWithCandidateLemmasIsTrue}), these conjectures are returned, which will be validated as sub-goals by recursively invoking \texttt{ProveRun}. Otherwise, \texttt{Prove} returns \texttt{False}.

\begin{figure}[t]
    \centering
\begin{algorithm}[H]
\small
\caption{The Prove Function}
\label{alg:prove-algorithm}
\SetKwProg{Fn}{Function}{:}{}
\SetKwFunction{ProveRun}{ProveRun}
\SetKwFunction{Preprocess}{Preprocess}
\SetKwFunction{Query}{LLMQuery}
\SetKwFunction{IsFiltered}{isFiltered}
\SetKwFunction{Prove}{Prove}
\SetKwFunction{Verify}{Verify}
\SetKwFunction{ProveParallel}{ProveParallel}
\tcp{Input: labeled SMTLIB2 file $P$, current prompt \emph{prompt};}
\tcp{Output:
\begin{tabular}{l}
 ({\tt True},$C$) if $P$ is proved with useful conjectures $C$; \\
({\tt False}, $\emptyset$) if conjectures cannot assist in proving $P$;
\end{tabular}
}
\Fn{\Prove{$P$, {\tt prompt}}}{
    $C\gets $ \Query{$P$, {\tt prompt}}\; \label{alg-line:query}
    \ForEach{$c \in C$}{ \label{algo:filtering}
        \If{\IsFiltered{$c$}}{ \label{alg-line:filtering}
            \Return ({\tt False}, $\emptyset$)\;
        }
    }
    \If{\Verify{$P$, $C$}}{ \label{alg-line:verifyWithCandidateLemmas}
        \Return ({\tt True}, $C$)\; \label{alg-line:verifyWithCandidateLemmasIsTrue}
    }
    
    \Return ({\tt False}, $\emptyset$)\;
}
\end{algorithm}
\vspace{-5mm}
\end{figure}

\smallskip\noindent
\textbf{The \texttt{isFiltered} function}. 
The filtering stage, implemented via \texttt{isFiltered}, 
is designed to quickly identify incorrect and useless conjectures.
An LLM-generated conjecture $c$ is filtered out by
\texttt{isFiltered}, i.e., \texttt{isFiltered}$(c)$ returns
{\tt True}, if (i)
$c$ has syntactic errors checked by the parser of an SMT solver
(e.g., cvc5), 
or (ii) $c$ is identical to the proof goal $P$ compared by 
syntactic equivalence checking, 
or (iii) $c$ is inconsistent with the axiom $A$, i.e.,
$A\wedge c$ is unsatisfiable, via SMT solving.
We check the inconsistency of $A$ and $c$ instead of the invalidity of $A \rightarrow c$ (equivalently, the satisfiability of $A \wedge \neg c$) 
because: $A\wedge c$ 
usually contains only the universal quantifier $\forall$ while $A\wedge \neg c$ contains both the universal quantifier $\forall$
and existential quantifier $\exists$, thus $A\wedge c$ is easier to be checked; moreover the unsatisfiability of $A\wedge c$ entails 
the satisfiability of $A\wedge \neg c$, thus $c$ cannot be 
a lemma under the axiom $A$.

\smallskip\noindent
\textbf{The \texttt{Verify} function}. 
The validation stage implemented via \texttt{Verify} is used to check whether the LLM-generated conjectures $C$ are \textit{useful} for proving the goal.
It is done by checking the unsatisfiability 
of $\{A, \neg P\} \cup C$, which is equivalent to checking $A \wedge \bigwedge_{c \in C} c\rightarrow P$. If the unsatisfiability 
of $\{A, \neg P\} \cup C$ can be proved by SMT solving, 
\texttt{Verify} returns {\tt True}, otherwise {\tt False}.

\subsection{Prompt Strategies for Lemma Generation}
\label{ssec:prompt}

To effectively elicit LLMs to generate useful lemmas, we design two prompt strategies, each of which consists of three parts: \texttt{Task Description} (explaining the SMTLIB2 format and the task), \texttt{Chain of Thoughts} (providing the reasoning strategy), and \texttt{Output Format} (specifying how to format the output).

\smallskip
\noindent
{\bf Strategy 1: Equational Reasoning.}
This strategy guides LLMs to perform step-by-step, human-like equational reasoning. Concretely, for a proof goal of the form $\forall x \forall \vec{y}.\ f(x, \vec{y}) = g(x, \vec{y})$, where $f$ is inductively defined over the variable $x$, the prompt instructs the LLM to: (i) identify the inductive definition of $f$ from the labeled SMTLIB2 file; (ii) determine whether the base case requires auxiliary lemmas and, if it can be proved by the solver's build-in simple induction, proceed directly to the inductive case;  (iii) for the inductive case $\forall x' \forall \vec{y}.\  f(t(x'), \vec{y}) = g(t(x'), \vec{y})$, where $x'$ is the variable used in the inductive hypothesis (e.g., $\forall x' \forall y.\ {\tt mult}({\tt succ}(x'), y) = {\tt mult}(y, {\tt succ}(x'))$), transform the left-hand side $f(t(x'), \vec{y})$ step-by-step using known premises (axioms and the inductive hypothesis); 
and (iv) when a step cannot be derived directly from known premises, generate it as a conjecture.

\begin{example}
To prove $\forall x, y \in \texttt{Nat}. \ \texttt{mult}(x, y) = \texttt{mult}(y, x)$ (i.e., the property in Listing~\ref{lst:adt-nat-example}), using Strategy 1, an LLM may reason as:
\begin{equation*} 
\label{eq:prompt-llm-in-nat-propery} 
\begin{aligned} 
\texttt{consider the inductive case:} \\[-3pt]
\texttt{mult(succ(x'), y))}
& =:| \ \texttt{axiom of mult} \ |: \\[-3pt]
\texttt{plus(mult(x', y), y)} 
& =:|  \ \texttt{inductive hypothesis} \ |: \\[-3pt]
\colorbox{yellow!30}{\texttt{plus(mult(y, x'), y)}}
&\colorbox{yellow!30}{=:| \ \texttt{unknown conjecture} \ |:} \\[-3pt]
\colorbox{yellow!30}{$\texttt{mult(y, succ(x'))}$}
\end{aligned}
\end{equation*}
The last highlighted step cannot be directly derived from known premises, so the following conjecture is generated by the LLM:
\begin{equation*}
    \forall \ x, y \in \texttt{Nat}. \ \texttt{plus}(\texttt{mult}(y, x), y) = \texttt{mult}(y, \texttt{succ}(x)).
\end{equation*}
Though this conjecture suffices to prove the property, it cannot be directly proved via SMT solving (e.g., cvc5 with 1200s time limit). Thus, it will be regarded as a sub-goal on which Strategy 1 is applied again, finally resulting in the following useful conjecture:
\begin{equation}
\begin{aligned}
\label{eq:prompt1-l1}
    \forall \ x, y \in \texttt{Nat}. \ &
    \texttt{plus}(\texttt{plus}(\texttt{mult}(y, x), x), \texttt{succ}(y)) = \\
   & \texttt{plus}(\texttt{plus}(\texttt{mult}(y, x), y), \texttt{succ}(x)).
     \end{aligned}
\end{equation}
\end{example}

\smallskip
\noindent
{\bf Strategy 2: Term Rewriting and Generalization.}
Unlike Strategy 1, this strategy does not instruct LLMs to think step-by-step, but provides common ideas in lemma generation, encouraging them to generate lemmas more flexibly and simplify the proof goal $P$. Concretely, the prompt instructs the LLM to: (i) generate basic axioms; (ii) strengthen the conclusion, i.e., 
find a stronger lemma that is easier to prove; (iii) identify
a common term that appears on both sides of the proof goal $P$ and 
replace it with a fresh variable, leading to a simplified proof goal $P'$; (iv) if no common terms exist, attempt to rewrite terms in $P$ using function definitions to have a common term; and (v) if necessary (e.g., $P'$ is not enough to prove $P$), generate bridging lemmas between $P'$ and $P$ as conjectures such as $P'\rightarrow L$ and  $L\rightarrow P$.

\begin{example} 
Suppose the LLM fails to generate 
useful lemmas for proving the sub-goal in Equation~(\ref{eq:prompt1-l1}), via Strategy 1. Then, Strategy 2 is used to guide the LLM, which identifies the common term ${\tt mult}(y, x)$, rewrites the sub-goal, and generates the following simplified sub-goal:
\begin{equation*}
    \forall \ t, x, y \in \texttt{Nat}. \ \texttt{plus}(\texttt{plus}(t, x), \texttt{succ}(y)) = \texttt{plus}(\texttt{plus}(t, y), \texttt{succ}(x)).
\end{equation*}
This simplified sub-goal can be directly proved via SMT solving,
thus we can complete the overall proof of the property.
\toggle{A further example for illustrating Strategy 2 is given in Appendix~\ref{sec:prompt2Example}.}
{A further example of Strategy 2 is given in the full version of the paper.}
\end{example}

\section{Evaluation}
\label{sec:evaluation}

We implement our approach in a tool, named \tool.
We evaluate \tool to answer the following research questions (\textbf{RQs}):
\begin{description}
\item[RQ1.] How effective is our approach at solving constraints involving inductive definitions compared with state-of-the-art solvers?
\item[RQ2.] What is the contribution of our prompt and filtering designs to the overall performance?
\item[RQ3.] Is our approach robust over different LLM models and sampling temperatures that controls 
creativity and diversity of the output?
\end{description}

\noindent\textbf{Benchmarks.}
We collected four inductive reasoning benchmarks from prior work:
StandardDT, StandardDTLIA, Autoproof, and IndBen,
which are commonly used in 
literature~\cite{ReynoldsK15,SunJFJCX24,KurashigeJGBNIJ24,AngelisFPP20,GovindShohamGurfinkel2022}. 
\begin{itemize}[topsep=0.2em,leftmargin=*]
\item StandardDT~\cite{ReynoldsK15} comprises 241 proof tasks in the ADT theory
which was collected from 
test suites of IsaPlanner~\cite{isaplanner}, CLAM~\cite{Ireland96} and HipSpec~\cite{ClaessenJRS13}, and verification conditions in functional program verification in Leon~\cite{BlancKKS13}. It is often regarded as a standard benchmark for automated inductive reasoning. 
    \item StandardDTLIA~\cite{AngelisFPP20}  
contains 168 proof tasks which were derived from StandardDT by converting the \texttt{Nat} type and all the related operations to \texttt{Int} in the LIA theory, covering both ADT and LIA theories.
    \item AutoproofBM~\cite{SunJFJCX24}
    comprises 141 proof tasks, which use more complex ADT and LIA definitions and are relatively more challenging to solve.
    \item  IndBen~\cite{HajduHKSV20} provides 3,396
    proof tasks among which we select 156 representative tasks,
    one task per group, 
    as the tasks in the same group define the same definitions and property (differing only in syntax).  
    The 156 proof tasks cover common ADT theories such as \texttt{List}, \texttt{Nat} and \texttt{Tree}.
\end{itemize}
Overall, we collected 706 proof tasks as our subject.

\smallskip
\noindent\textbf{Experimental Setup.}
To answer \textbf{RQ1}, we compare \tool with cvc5, Vampire and Racer in terms of the number of solved proof tasks.

\begin{itemize}[topsep=0.2em,leftmargin=*]
\item cvc5~\cite{BarbosaBBKLMMMN22} is one of the leading SMT solvers. 
As its performance may vary with options and versions~\cite{SunJFJCX24,KurashigeJGBNIJ24},
we run cvc5 (with three option configurations) and cvc4 (with the default inductive reasoning setting) in parallel
\toggle{(cf.\ Appendix~\ref{sec:cvc-options}),}{(cf.\ the full version of the paper for detail options),}
which serves as the baseline for cvc5 and acts as the backend SMT solver in our workflow (for clarity, it is referred to directly as cvc5).
\item Vampire~\cite{KovacsV13,HajduKRV22} is a leading automated first-order logic prover based on superposition calculus. It supports a fragment of the SMTLIB2 inputs. We run Vampire under the portfolio mode, which runs multiple proving strategies in parallel.  
 \item Racer~\cite{GovindShohamGurfinkel2022} extends 
 the CHC solver Spacer~\cite{KomuravelliGC16} to better support ADTs and RDFs. As it only supports inputs in CHCs (not general SMTLIB2) form, 
 we only compare with it using its provided benchmarks.
\end{itemize}

To answer \textbf{RQ2}, we conduct an ablation study.
To answer \textbf{RQ3}, we evaluate \tool using four different LLMs: DeepSeek-v3.2 (DeepSeek), Qwen3-235B-instruct (Qwen), Gemini 2.5 flash (Gemini) and GPT-5 and vary the sampling temperature.
Note that we use Qwen as the default LLM unless stated explicitly otherwise,
as a trade-off between the cost and performance.\footnote{The price per M input/output tokens: DeepSeek-v3.2 (\$0.27/\$0.40), Qwen3-235B (\$0.072/\$0.464), Gemini 2.5 flash (\$0.30/\$2.50), GPT-5 (\$1.25/\$10.00).} 

All experiments were conducted on a machine
with an Intel(R) Xeon(R) Platinum 8368Q CPU (76 cores, 2.60 GHz), 1024 GiB RAM and Ubuntu 22.04.5 LTS
(1024 GiB RAM is the memory for evaluation instead of consumption). 
The solving of each proof task is limited to 1200 seconds (wall time) unless stated explicitly otherwise. 
For \tool, we use both prompt strategies described in Section~\ref{ssec:prompt}, each invoked for up to 3 iterations as in Algorithm~\ref{alg:overall-algorithm} (line~\ref{alg-line:prompt-strateges-begin} and line~\ref{alg-line:max-iter-numbers}).
The backend solver is given 60 seconds for initial checking (Algorithm~\ref{alg:overall-algorithm}, line~\ref{alg-line:initialcheck}) and for each invocation of the \texttt{Verify} function to verify candidate conjectures (Algorithm~\ref{alg:prove-algorithm}, line~\ref{alg-line:verifyWithCandidateLemmas}).
The filtering checking (Algorithm~\ref{alg:prove-algorithm}, line~\ref{alg-line:filtering}) is given 1 second per conjecture.
At most 4 threads are used to verify sub-goals concurrently (Algorithm~\ref{alg:overall-algorithm}, line~\ref{alg-line:parallel-begin}).
We set the hyperparameter top\_p=0.9 and sampling temperature=0.9 by default according to~\cite{abs-2506-07295}. For cvc5, we run 4 threads in parallel, one thread per option configuration. 
Other tools use their default thread configurations without additional parallelization settings.  

\subsection{RQ1. Effectiveness} 

The results are reported in Table~\ref{tab:rq1-effectiveness},
including the number of tasks solved by each tool with the time limits 1200s and 360s, as well as the average solving time of the solved tasks. Recall that Racer cannot directly verify StandardDT, AutoProofBM and IndBen (hence is marked by N/A). 
For reference, a full run of \tool over all benchmarks consumed approximately 11.5M tokens using Qwen, corresponding to an estimated cost of about \$4.

Overall, \tool solved significantly more tasks than the baselines, regardless of the time limit. For instance, under 1200s, \tool solved 232 more tasks than cvc5 and 182 more tasks than Vampire. 
This confirms the effectiveness of our approach. 
On StandardDTLIA, \tool also solved 94 more tasks than Racer. 
On IndBen, Vampire solved more tasks than \tool, but we will see later in Table~\ref{tab:ablation-prompts} that \tool surpasses 
Vampire when more powerful LLMs (Gemini and GPT-5) are used instead of Qwen.

It is not surprising that \tool's solving time is higher than the traditional logic-based approaches, because it recursively queries the LLM and invokes the SMT solver.
Arguably, the time remains acceptable for verification tasks (on average about 100s), especially when producing a correct result is more critical. 

\begin{table}[t]\footnotesize
 \centering\setlength{\tabcolsep}{2pt}  
\caption{Comparison with baseline SMT solvers}\centering
 \scalebox{0.9}{ 
 \begin{tabular}{l|c|cc|cc|cc|cc}
\hline
\multirow{2}{*}{\textbf{Benchmark}} & \multirow{2}{*}{\textbf{Total}} & \multicolumn{2}{c|}{\tool} & \multicolumn{2}{c|}{cvc5} & \multicolumn{2}{c|}{Vampire} & \multicolumn{2}{c}{Racer} \\
\cline{3-10}
&  & <1200s & <360s & <1200s & <360s & <1200s & <360s & <1200s & <360s \\
\hline
StandardDT & 241 & \textbf{212} & \textbf{200} & 150 & 150 & 160 & 160 & N/A & N/A \\
StandardDTLIA & 168 & \textbf{134} & \textbf{127} & 76 & 76 & 58 & 56 & 40 & 40 \\
AutoProofBM & 141 & \textbf{65} & \textbf{61} & 34 & 34 & 3 & 3 & N/A & N/A \\
IndBen & 156 & 114 & 104 & 34 & 34 & \textbf{122} & \textbf{118} & N/A & N/A \\
\hline
\textbf{Total} & 706 & \textbf{525} & \textbf{492} & 293 & 293 & 343 & 337 & 40 & 40 \\
\hline
\textbf{Avg time (s)} &  & 97.30 & 58.66 & \textbf{3.38} & \textbf{3.38} & 19.39 & 9.26 & 6.09 & 6.09 \\
\hline
\end{tabular}}
\label{tab:rq1-effectiveness}
\vspace{-2mm}
\end{table}



We note that the recent tools AutoProof~\cite{SunJFJCX24} and CCLemma~\cite{KurashigeJGBNIJ24} have reported promising results on lemma generation for inductive reasoning. However, these tools 
support neither SMTLIB2 nor CHC format inputs.
As an \emph{indirect} comparison using the results reported in~\cite{SunJFJCX24}, 
AutoProof solved 161 tasks 
while \tool solved 180 tasks using their benchmarks.


\subsection{RQ2. Ablation Study}
To study the contribution of our prompt and filtering designs, we compare our prompt strategies with the naive prompt strategy (cf.~Listing~\ref{lst:naive-prompts})
and evaluate \tool with and without the filter. 


\begin{table}[t]
\centering \setlength{\tabcolsep}{3pt}   
\caption{Results of \tool using different LLMs and naive prompt strategy}
\label{tab:ablation-prompts}
\scalebox{0.9}{
\begin{tabular}{l|c|c|c|c|c|c|c|c}
\hline
\multirow{2}{*}{\textbf{Benchmark}} & \multicolumn{2}{c|}{Qwen} & \multicolumn{2}{c|}{DeepSeek} & \multicolumn{2}{c|}{Gemini} & \multicolumn{2}{c}{GPT-5} \\
\cline{2-9}
 & \tool & Naive & \tool & Naive & \tool & Naive & \tool & Naive \\
\hline
StandardDT & \textbf{212} & 160 & \textbf{206} & 153 & \textbf{206} & 155 & \textbf{212} & 180 \\
StandardDTLIA & \textbf{134} & 88 & \textbf{131} & 90 & \textbf{119} & 91 & \textbf{144} & 106 \\
AutoProofBM & \textbf{65} & 42 & \textbf{63} & 43 & \textbf{68} & 49 & \textbf{67} & 46 \\
IndBen & \textbf{114} & 89 & \textbf{121} & 79 & \textbf{126} & 76 & \textbf{129} & 118 \\
\hline
\textbf{Total} & \textbf{525} & 379 & \textbf{521} & 365 & \textbf{514} & 371 & \textbf{552} & 450 \\
\hline
\end{tabular}}
\vspace{-4mm}
\end{table}

\smallskip
\noindent{\bf Prompt Design.}
The results are reported in Table~\ref{tab:ablation-prompts}
using four different LLMs: Qwen, DeepSeek, Gemini and GPT-5. Here, \tool uses the prompt strategies presented in Section~\ref{ssec:prompt}, while the
naive one (cf.\ Listing~\ref{lst:naive-prompts}) provides only a simple task environment description, requests lemma generation, and specifies output format. For fairness, the naive prompt is invoked up to 6 times, matching the others (2 prompts$\times$3 times/prompt).
\tool with our prompt strategies significantly outperforms
the naive one, across all LLM models and benchmarks.



\smallskip
\noindent{\bf Filtering Design.}
The results are reported in Table~\ref{tab:filtering-time}. 
Because solving outcomes depend heavily on the quality of LLM outputs, we run each setting (with and without the filter) three times (R1, R2 and R3) and report the \hide{results of each run to assess the overall effect} average results (Avg). For a fair comparison, the average time is computed over all solved tasks. 
For per-task time comparison across all four benchmarks, we provide scatter plots in 
\toggle{Appendix~\ref{sec:filtering-time-comparison},} 
{the full version of the paper,} visualizing the solving time differences for each proof task.

In general, with the filter, \tool solves more tasks (520.7 vs.\ 513.7 in total over three runs), indicating that the filtering design is effective. In many cases, without the filter (w.o.\ filter), incorrect or useless conjectures consume substantial solving time and can lead to 
time out. In contrast, with the filter (w.\ filter), some iterations are discarded early, saving time and allowing those instances to be solved successfully.

In detail, while the filter can reduce the solving time on the three benchmarks (i.e., StandardDTLIA, AutoProofBM and IndBen), StandardDT remains an exception. It may be because StandardDT contains relatively simple ADT definitions (primarily common types such as \texttt{List}, \texttt{Nat}, and \texttt{Tree}) for which LLMs can generate correct conjectures quickly, so that filtering provides little benefit but incurs additional overhead.
In addition, the last row of Table~\ref{tab:filtering-time} reports the total tokens (in millions) consumed across the six runs. Overall, enabling the filter results in lower token usage compared to the configuration without the filter, indicating improved cost efficiency.

\begin{table}[t]\setlength{\tabcolsep}{4pt}\small
\caption{Results of \tool with and without filter over three runs}\centering
\scalebox{0.86}{
\begin{tabular}{l|cccc|cccc|cc}
\hline
\multirow{2}{*}{\textbf{Benchmark}} & \multicolumn{4}{c|}{W. filter} & \multicolumn{4}{c|}{W.o. filter} & \multicolumn{2}{c}{Avg time (s)} \\
\cline{2-11}
 & R1 & R2 & R3 & Avg & R1 & R2 & R3 & Avg & W. filter & W.o. filter \\
\hline
StandardDT & 212 & 211 & 207 & \textbf{210} & 204 & 203 & 209 & 205.3 & 83.91 & \textbf{67.26} \\
StandardDTLIA & 134 & 137 & 135 & 135.3 & 134 & 139 & 134 & \textbf{135.7} & \textbf{97.53} & 112.05 \\
AutoProofBM & 65 & 61 & 64 & \textbf{63.3} & 60 & 64 & 64 & 62.7 & \textbf{99.10} & 103.61 \\
IndBen & 114 & 111 & 111 & \textbf{112} & 113 & 106 & 111 & 110 & \textbf{168.92} & 178.64 \\
\hline
\textbf{Total} & 525 & 520 & 517 & \textbf{520.7} & 511 & 512 & 518 & 513.7 & 107.59 & \textbf{107.37} \\
\hline
Tokens (M) & 11.50 & 12.08 & 11.57 & 11.72 & 13.37 & 11.75 & 12.49 & 12.54 & N/A & N/A \\
\hline
\end{tabular}}
\label{tab:filtering-time}
\vspace{-4mm}
\end{table}


\subsection{RQ3. Robustness of \tool} 
To evaluate the robustness of {\tool} across different LLMs, we 
consider DeepSeek, Gemini and GPT-5, besides Qwen. The experimental results are presented in Table~\ref{tab:ablation-prompts}, from which we observe that {\tool} consistently improves cvc5, Vampire and Racer, regardless of the LLM model used. 
Moreover, to assess robustness under different samplings, we set the temperature of Qwen to $0.1$, $0.5$, $0.9$ and $1.3$, respectively. 
We ran three independent runs for each temperature to account for the randomness of LLM outputs. 
Table~\ref{tab:temperature} reports the per-run counts. For each temperature, we 
also report the range (max$-$min over the three runs) and the standard deviation (std) of the total number of solved tasks to quantify variability. The range 
is at most 9 (e.g., 518--527 for $T=1.3$), and the std remains below 5 in all cases, indicating low variance across runs.
We observe that as the temperature increases, the variability tends to increase slightly (e.g., the std grows from 3.2 at $T=0.1$ to 4.7 at $T=1.3$, and the range increases from 6 to 9), which aligns with the expectation that higher temperatures introduce greater randomness into LLM sampling. 
In summary, the number of solved tasks varies only slightly across runs and temperatures.
These results suggest that the inherent randomness of LLMs and the sampling temperature have a minor impact on {\tool}'s performance, demonstrating its robustness over different runs and temperatures.  

\begin{table}[t]
\centering\setlength{\tabcolsep}{4pt}\small
\caption{Results of \tool with different sampling temperatures}
\label{tab:temperature}
\scalebox{0.9}{
\begin{tabular}{l|ccc|ccc|ccc|ccc}
\hline
\multirow{2}{*}{\textbf{Benchmark}} & \multicolumn{3}{c|}{T = 0.1} & \multicolumn{3}{c|}{T = 0.5} & \multicolumn{3}{c|}{T = 0.9} & \multicolumn{3}{c}{T = 1.3} \\
\cline{2-13}
 & R1 & R2 & R3 & R1 & R2 & R3 & R1 & R2 & R3 & R1 & R2 & R3 \\
\hline
StandardDT & 211 & 206 & 210 & 207 & 209 & 207 & \textbf{212} & 211 & 207 & 208 & 209 & 210 \\
StandardDTLIA & 135 & 139 & 135 & 136 & 138 & 138 & 134 & 137 & 135 & \textbf{140} & 136 & 135 \\
AutoProof & 60 & 62 & 62 & 61 & 62 & 61 & \textbf{65} & 61 & 64 & 60 & 60 & \textbf{65} \\
IndBen & 114 & 108 & 114 & 110 & 107 & 113 & 114 & 111 & 111 & \textbf{119} & 113 & 112 \\
\hline
\textbf{Total} & 520 & 515 & 521 & 514 & 516 & 519 & 525 & 520 & 517 & \textbf{527} & 518 & 522 \\
\hline
\multicolumn{1}{l|}{\footnotesize Range \ Std} & \multicolumn{3}{c|}{6 \quad 3.2} & \multicolumn{3}{c|}{5 \quad 2.6} & \multicolumn{3}{c|}{8 \quad 4.2} & \multicolumn{3}{c}{9 \quad 4.7} \\
\hline
\end{tabular}}
\vspace{-3mm}
\end{table}


\begin{table}[t]\footnotesize
 \centering\setlength{\tabcolsep}{2pt}  
\caption{Comparison with baseline SMT solvers (Vampire backend)}\centering
 \scalebox{0.9}{ 
 \begin{tabular}{l|c|cc|cc|cc|cc}
\hline
\multirow{2}{*}{\textbf{Benchmark}} & \multirow{2}{*}{\textbf{Total}} & \multicolumn{2}{c|}{\tool-V} & \multicolumn{2}{c|}{cvc5} & \multicolumn{2}{c|}{Vampire} & \multicolumn{2}{c}{Racer} \\
\cline{3-10}
&  & <1200s & <360s & <1200s & <360s & <1200s & <360s & <1200s & <360s \\
\hline
StandardDT & 241 & \textbf{214} & \textbf{206} & 150 & 150 & 160 & 160 & N/A & N/A \\
StandardDTLIA & 168 & \textbf{80} & 73 & 76 & \textbf{76} & 58 & 56 & 40 & 40 \\
AutoProofBM & 141 & 19 & 16 & \textbf{34} & \textbf{34} & 3 & 3 & N/A & N/A \\
IndBen & 156 & \textbf{140} & \textbf{140} & 34 & 34 & 122 & 118 & N/A & N/A \\
\hline
\textbf{Total} & 706 & \textbf{453} & \textbf{435} & 293 & 293 & 343 & 337 & 40 & 40 \\
\hline
\textbf{Avg time (s)} &  & 59.89 & 38.74 & \textbf{3.38} & \textbf{3.38} & 19.39 & 9.26 & 6.09 & 6.09 \\
\hline
\end{tabular}}
\label{tab:vampire-backend}
\vspace{-3mm}
\end{table}
We also replace cvc5 in our workflow with Vampire (denoted as \tool-V). 
Table~\ref{tab:vampire-backend} reports the results. Overall, \tool-V solved more proof tasks than each of the three baselines. Compared with Vampire, \tool-V shows improvements across all four benchmarks, demonstrating that our approach can enhance Vampire's performance and that \tool remains robust across different backend solvers. 
However, compared to cvc5, \tool-V underperforms in two cases. In particular, for StandardDTLIA mixed with ADT and LIA theories, Vampire has limited support, so \tool-V solved fewer tasks than cvc5 under the 360s limit. 
%
For AutoProofBM, 119 (out of 141) proof tasks involve ADT symbols defined in the SMTLIB2 standard. They are supported by cvc5, but not by Vampire. As a result, although our approach improves Vampire's performance, \tool-V still solved fewer tasks than cvc5 on this benchmark.

\section{Related Work} \label{sect:related}
Inductive reasoning is supported, with different degrees of human intervention, by many theorem provers. 
Existing works address automated inductive reasoning mainly from two perspectives.
(1) Incorporating \textit{induction schemas} into decision procedures that originally do not support induction. 
For instance, \cite{ReynoldsK15} enabled SMT solvers to automatically handle inductive definitions by integrating induction schemas into quantifier elimination, while works~\cite{Cruanes17,RegerV19,DBLP:journals/jar/EchenimP20,HajduHKV21,HozzovaKV21,HajduKRV22} extended the superposition calculus with induction rules.
(2) Generating auxiliary lemmas to enhance inductive reasoning in provers that support induction, such as automated theorem provers for functional languages, including IsaPlanner~\cite{isaplanner}, ACL2~\cite{kaufmann2013computer}, Zeno~\cite{SonnexDE12} and HipSpec~\cite{ClaessenJRS13}.
Below, we briefly review representative approaches on automated lemma generation.

\smallskip

\noindent{\bf Theory Exploration.} 
The main idea 
is to construct a pool of candidate lemmas from available symbols, 
typically by term enumeration or instantiating formula templates. 
The candidates are then filtered using heuristics. 
HipSpec~\cite{ClaessenJRS13} is a representative tool in this category, which uses counterexample and congruence closure to efficiently filter out invalid lemmas. CCLemma~\cite{KurashigeJGBNIJ24} borrows this pruning idea and proposes e-graph guided lemma discovery to make theory exploration more goal-directed. In addition, cvc5~\cite{ReynoldsK15} also adopts a similar idea: it enumerates conjectures in increasing size, and filters them using techniques based on active conjectures, equivalence-class construction, and ground-facts reasoning. ADTInd~\cite{YangFG19} employs syntax-guided templates to enumerate lemmas and implements a generalization mechanism for refining them.

\smallskip
\noindent{\bf Generalization.} Generalization techniques identify common sub-terms in the proof goal and replace them with fresh variables to find auxiliary lemmas that are simpler and can generalize the target property.
The main heuristic is to select the common sub-term for replacement by introducing new variables in positions where induction can potentially be applied. 
This approach has been adopted by modern theorem provers such as ACL2~\cite{kaufmann2013computer}. 
Zeno~\cite{SonnexDE12} applies the same common sub-term technique but combines it with counterexample searching to avoid over-generalization. 
Vampire~\cite{KovacsV13,HajduHKSV20} introduces an \textit{induction with generalization} inference rule within the superposition calculus framework. This rule can handle properties with multiple occurrences of the same induction term, and instantiates induction axioms with logically stronger variants of the property being proved.  
AutoProof~\cite{SunJFJCX24} differs from generalization-based approaches, but can be regarded as a goal-oriented method. It transforms the goal into \textit{induction-friendly forms} that guarantee effective use of inductive hypotheses. Moreover, it synthesizes lemmas as equations where one side matches a term in the goal, enabling systematic rewriting in proof. This directed lemma synthesis avoids enumerating useless lemmas and ensures progress toward provable sub-goals. 

\smallskip
\noindent{\bf CHC-Based Methods.} These approaches treat inductive definitions as transition systems and solve the problem by synthesizing inductive invariants. 
Several CHC solvers can handle ADTs, but reasoning about RDFs remains challenging due to the undecidability of the underlying logic. 
For instance, VeriMAP~\cite{AngelisFPP18a,AngelisFPP20} transforms CHCs with ADTs to CHCs over basic types, but the transformation is unsound for UNSAT answers. 
Racer~\cite{GovindShohamGurfinkel2022} addresses this challenge by approximating RDFs abstractions. It compiles RDFs to CHCs while preserving satisfiability 
and replaces RDFs with finite unfoldings parameterized by depth $k$. It implements an IC3-style algorithm, enabling automatic learning of inductive invariants over ADTs and RDFs. 

\smallskip
\noindent{\bf Neuro-Symbolic Methods.}
These approaches largely follow a generate-then-verify paradigm, in which neural models propose candidate lemmas, and symbolic tools are used to validate or refute them.
Several works apply this idea in deductive verification or ITPs, where LLMs assist users by proposing lemmas that are subsequently checked by a verifier or proof assistant (e.g., Dafny~\cite{SilvaMF24} and Verus~\cite{YangLMYCGHLLL0Z25}, and ITPs such as Lean4~\cite{VaramballyVSCYY25} and Isabelle~\cite{AlhessiEGFLJS25}).
These approaches primarily aim to enhance user productivity or guide interactive proof search, rather than enable fully automated reasoning.

Recent work~\cite{PeledKTV25} studies LLM-assisted invariant generation for hardware model checking, where LLMs propose candidate inductive invariants over hardware transition systems. Similarly, 
\cite{GauthierU25} learns induction predicates for inductive and arithmetic problems using traditional neural models. It focuses on conjectures over inductive definitions encoded as SMT formulas in integer theory, and is not designed as a general LLM-based framework.
AquaForte~\cite{LvDHJMZ26} applies LLMs to generate candidate instantiations for quantified uninterpreted functions in SMT solving, primarily targeting satisfiability checking. 
In contrast, our work focuses on automated proof generation by establishing the unsatisfiability of the negation of the target property.


\section{Conclusion} \label{sec:conclusion}
In this work, we have proposed a neuro-symbolic approach that leverages LLMs to generate auxiliary lemmas for inductive reasoning. Through a three-stage workflow, we synergistically integrate LLMs and constraint solvers, design prompt strategies to guide LLMs in generating high-quality candidate lemmas, and filter and validate them systematically. The experimental results on diverse benchmarks demonstrate the effectiveness of our approach. 

In the future, extending the approach to handle more general proof tasks and investigating other prompting and inference-time techniques, or agent-based methods, would be promising.

\begin{credits}
\subsubsection{\ackname} 
This work is supported by the Strategic Priority Research Program of the Chinese Academy of Sciences, Grant No.~XDA0320101. T. Chen is partially supported by overseas grants from the State Key Laboratory of Novel Software Technology, Nanjing University (KFKT2023A04,  KFKT2025A05).
\end{credits}


\bibliographystyle{splncs04}
\bibliography{submission-ref}
%
\toggle{
\appendix
\section{A Failed Case Study of cvc5}
\label{sec:examplecvc5}

We present a failed verification attempt using cvc5, the representative state-of-the-art solver, for proving the \code{property} in Listing~\ref{lst:adt-nat-example}.\footnote{
The actual inputs given to cvc5 are in standard SMTLIB2 format.}
Its lemma generation module, integrated in the 
DPLL(T) framework, starts by enumerating terms up to a given maximum size and combining them into conjectures such as  
\begin{equation*} 
\begin{aligned}
& \forall \ x, y \in \texttt{Nat}. \  x = \texttt{mult}(x, y),  \qquad
\forall \ x \in \texttt{Nat}. \  x = \texttt{succ}(x), \\
& \forall \ x, y \in \texttt{Nat}. \  y = \texttt{mult}(x, y),\qquad
\forall \ x, y \in \texttt{Nat}. \  y = \texttt{succ}(x), \\
& \forall \ x, y \in \texttt{Nat}. \  \texttt{zero} = \texttt{mult}(x, y), \quad
\forall \ x \in \texttt{Nat}. \  \texttt{zero} = \texttt{succ}(x), \cdots 
\end{aligned}
\end{equation*}
among which trivially invalid and useless ones
are heuristically and quickly filtered out.
The remaining conjectures are considered as candidate lemmas (e.g., the following three lemmas),
which are further verified by cvc5 and used to prove 
the original property once the candidate lemmas themselves are proved.  
\begin{equation*} 
\begin{aligned}
& \forall \ x, y \in \texttt{Nat}. \  x = \texttt{mult}(\texttt{succ}(\texttt{zero}), x), \\
& \forall \ x, y \in \texttt{Nat}. \  y = \texttt{plus}(\texttt{mult}(x, \texttt{zero}), y), \\
& \forall \ x \in \texttt{Nat}. \  \texttt{succ}(x) = \texttt{plus}(\texttt{succ}(\texttt{zero}), x)
\end{aligned}
\end{equation*}

In this example, the candidate lemmas fail to assist in proving the original property. As the size of the enumerated terms increases, cvc5 spends increasing amounts of time searching for candidate lemmas that can pass the filter. Even when some lemmas do pass, they may still fail to assist the verification. Consequently, cvc5 gets stuck and is unable to solve this proof task within the 1200s time limit. This example shows that the technique used by cvc5 is effective only for proof tasks that can be solved with relatively simple lemmas. 

\section{An Example of Prompt Strategy 2}
\label{sec:prompt2Example}
Here we present a further example to show how Prompt Strategy 2 works.
As shown in Listing~\ref{lst:prompt2ExampleApp}, we define \code{Nat}, \code{plus}, and \code{leq}, where \code{Nat} and \code{plus} are as defined in~\ref{lst:adt-nat-example}, and \code{leq} is a recursively defined function representing the less-than-or-equal relation. 
The \code{property} to be verified is at line~\ref{lst:property-app}, which states that for any natural number $x$, $x + x \leq (x + (x + x)) + x$. 
Due to the complex recursive definitions, cvc5 cannot prove the property within the 1200s time limit.
\begin{lstlisting}[language=SMT, style=SMT,
  label=lst:prompt2ExampleApp,
  caption={An example of prompt strategy 2}]
datatypes Nat := zero | succ(n: Nat) (*\label{lst:adt-nat-app}*)
fun plus(Nat, Nat): Nat { (*\label{lst:fun-plus-begin-app}*)
  (*$\forall$*)y(*$\in$*)Nat.plus(zero,y) = y
  (*$\forall$*)x(*$\in$*)Nat,y(*$\in$*)Nat.plus(succ(x),y) = succ(plus(x,y))
} (*\label{lst:fun-plus-end-app}*)
fun leq(Nat, Nat): Bool { (*\label{lst:fun-leq-begin-app}*)
  (*$\forall$*)y(*$\in$*)Nat.leq(zero,y)
  (*$\forall$*)x(*$\in$*)Nat,y(*$\in$*)Nat.leq(succ(x),succ(y)) = leq(x,y)
} (*\label{lst:fun-leq-end-app}*)
property: 
  (*$\forall$*)x(*$\in$*)Nat.leq(plus(x,x),plus(plus(x,plus(x,x)),x)) (*\label{lst:property-app}*)
\end{lstlisting}

With the Strategy 2, the LLM generates the following two conjectures:
\begin{equation}
\label{eq:lemma1-app}
    \forall x \in \texttt{Nat}. \ \texttt{leq}(\texttt{plus}(x, x), \texttt{plus}(x, \texttt{plus}(x, x)))
\end{equation}
\begin{equation}
\label{eq:lemma2-app}
    \forall a, b, c \in \texttt{Nat}. \ \texttt{leq}(a, b) \rightarrow \texttt{leq}(a, \texttt{plus}(b, c))
\end{equation}

With these two conjectures, cvc5 can successfully prove the original property. And each conjectures can be proved finally in our workflow (the first conjecture needs further auxiliary lemma, and the second conjecture can be proved directly by SMT solving). 

These two conjectures can be viewed as generalizing and simplifying the original property. 
By applying Equation~\ref{eq:lemma2-app} to Equation~\ref{eq:lemma1-app} with replacing $a$ to $\texttt{plus}(x, x)$, $b$ to $\texttt{plus}(x, \texttt{plus}(x, x))$, and $c$ to $x$, we obtain:
\begin{equation*}
    \forall x. \ \texttt{leq}(\texttt{plus}(x, x), \texttt{plus}(\texttt{plus}(x, \texttt{plus}(x, x)), x))
\end{equation*}
which proves the original property at line~\ref{lst:property-app}.

\section{Option Configurations of cvc5 and cvc4}
\label{sec:cvc-options}
We use the following three option configurations for cvc5
and one option configurations for cvc4:
\begin{itemize}[topsep=0.2em,leftmargin=*]
\item cvc5 --full-saturate-quant
  \item cvc5 --full-saturate-quant, --quant-ind, --conjecture-gen
  \item cvc5 --full-saturate-quant, --quant-ind, --conjecture-gen, --no-e-matching
  \item cvc4 --quant-ind, --quant-cf, --conjecture-gen, --full-saturate-quant, --lang=smt2.6
\end{itemize}
Detailed description of these options refer to~\cite{cvc5Document}

\section{Detail Results of Filtering}
\label{sec:filtering-time-comparison}

\begin{figure}[t]
\centering
\begin{minipage}{0.48\textwidth}
\centering
\includegraphics[width=\textwidth]{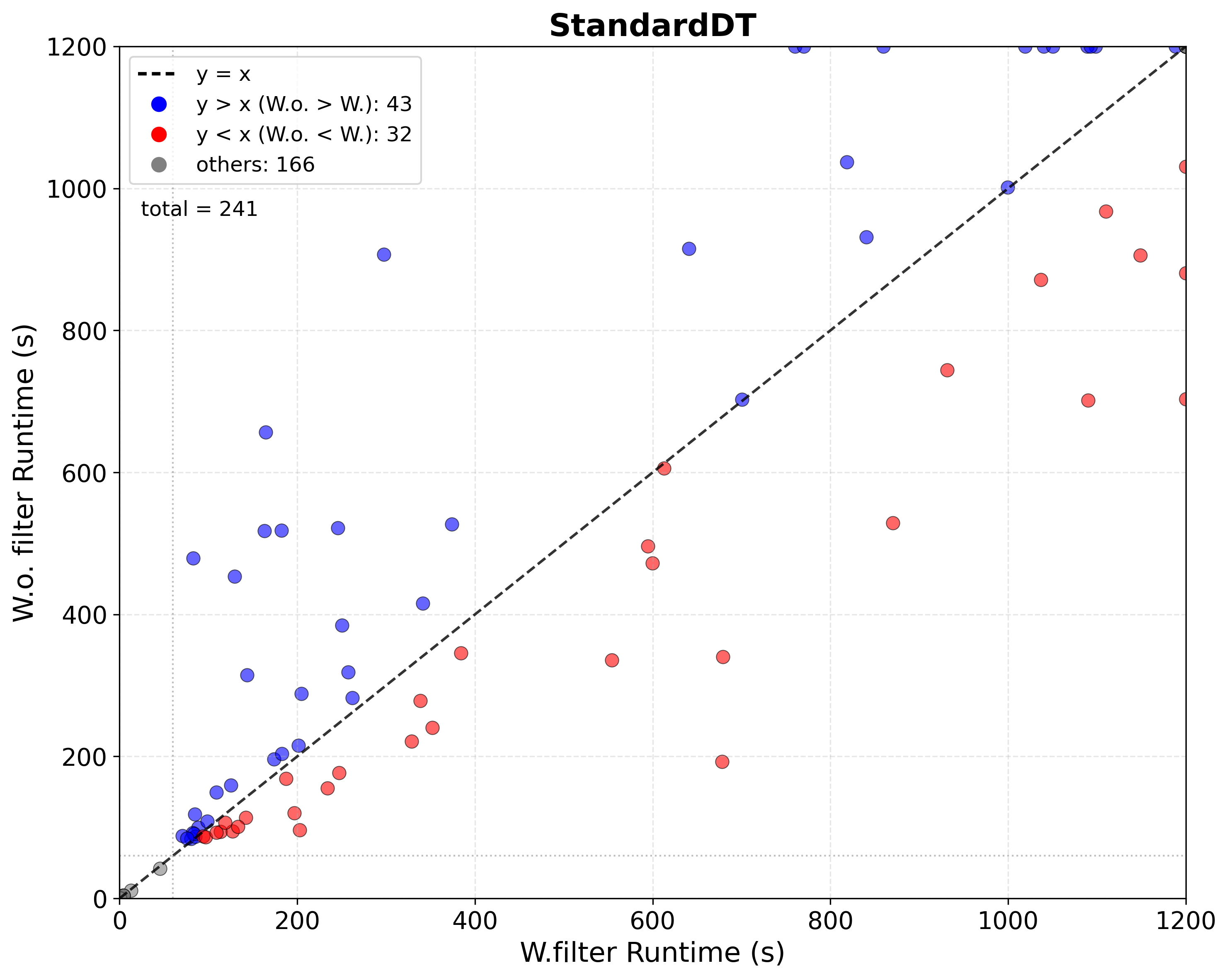}
\caption{StandardDT benchmark}
\label{fig:time-comparison-standarddt}
\end{minipage}
\hfill
\begin{minipage}{0.48\textwidth}
\centering
\includegraphics[width=\textwidth]{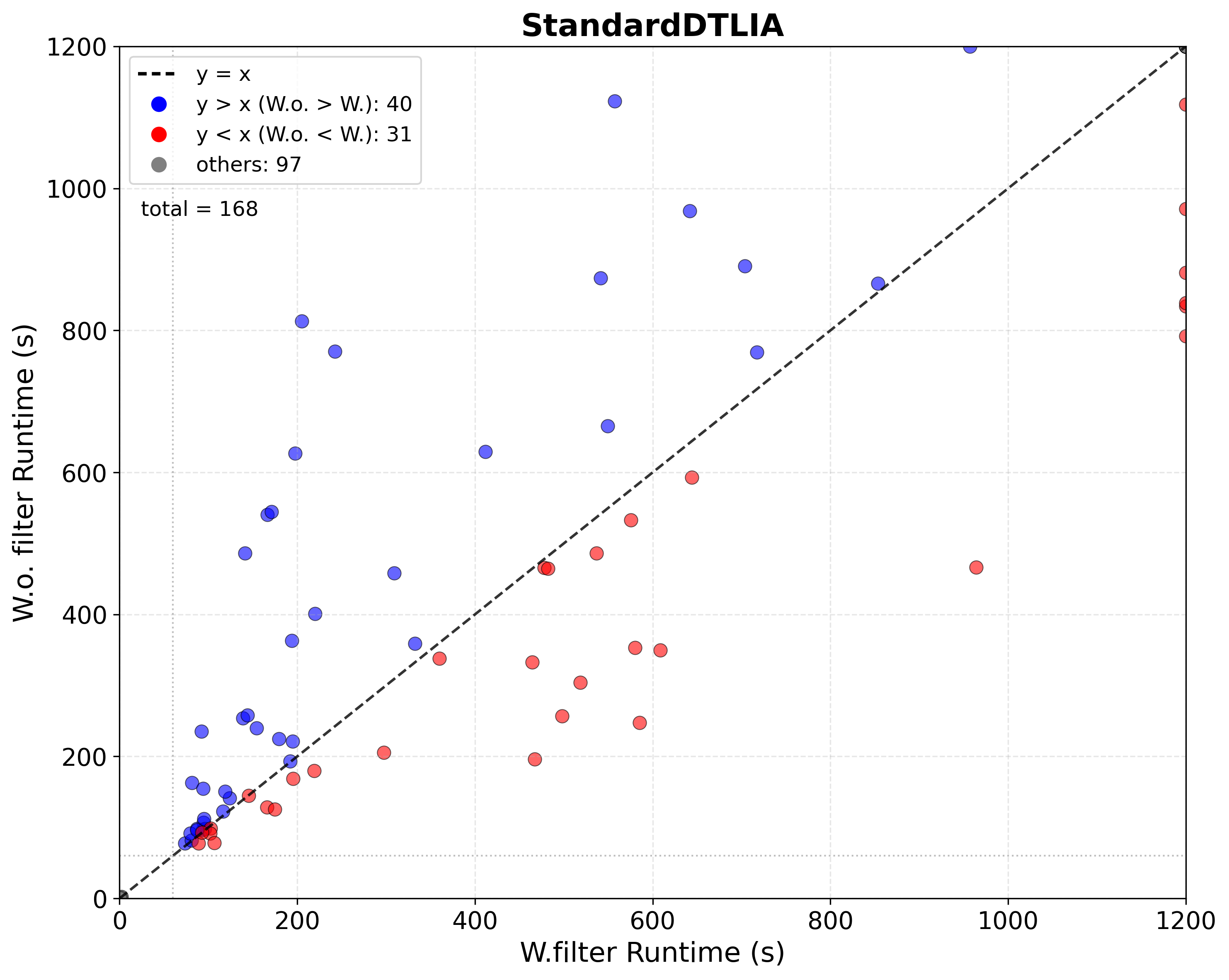}
\caption{StandardDTLIA benchmark}
\label{fig:time-comparison-standarddtlia}
\end{minipage}
\begin{minipage}{0.48\textwidth}
\centering
\includegraphics[width=\textwidth]{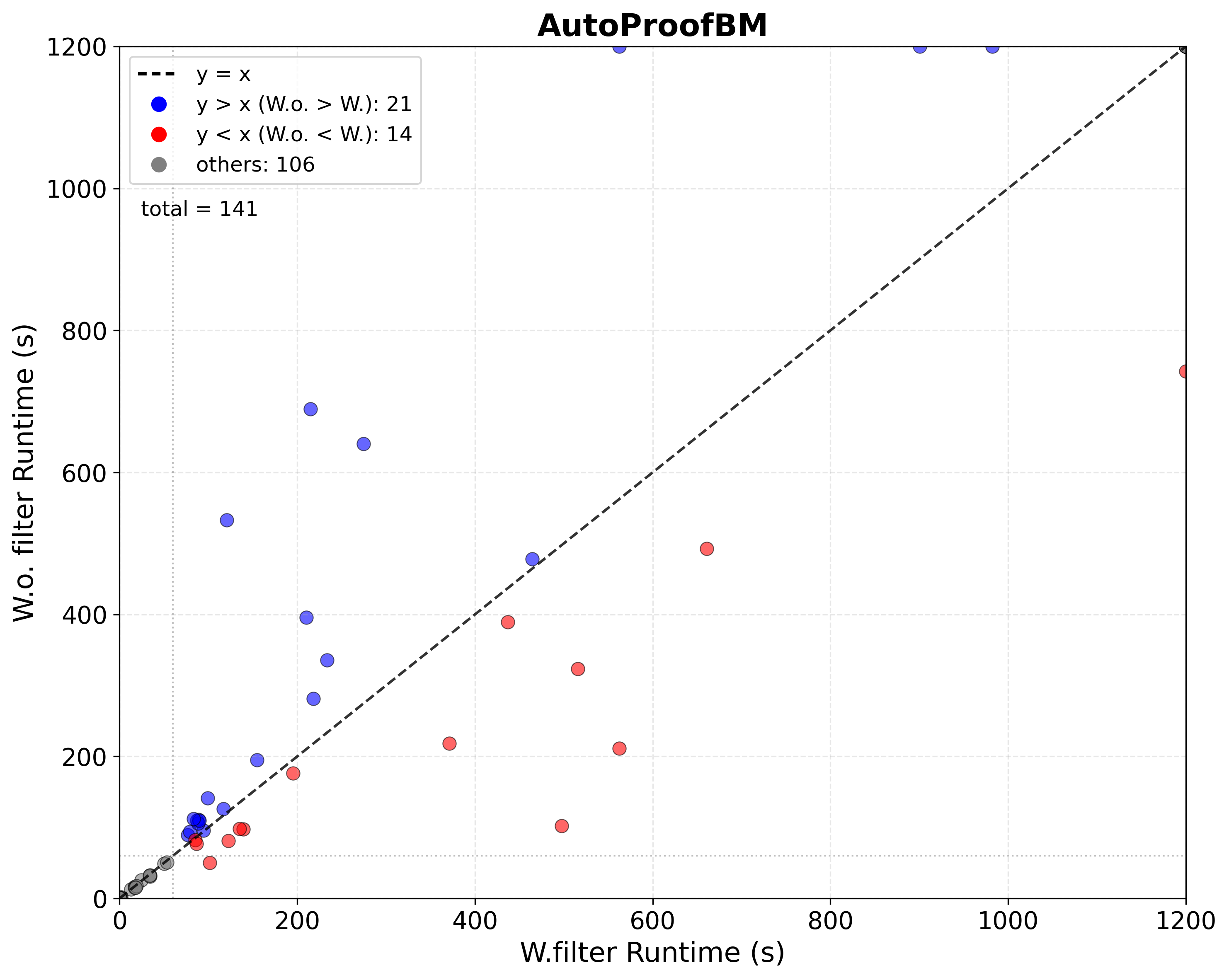}
\caption{AutoProofBM benchmark}
\label{fig:time-comparison-autoproof}
\end{minipage}
\hfill
\begin{minipage}{0.48\textwidth}
\centering
\includegraphics[width=\textwidth]{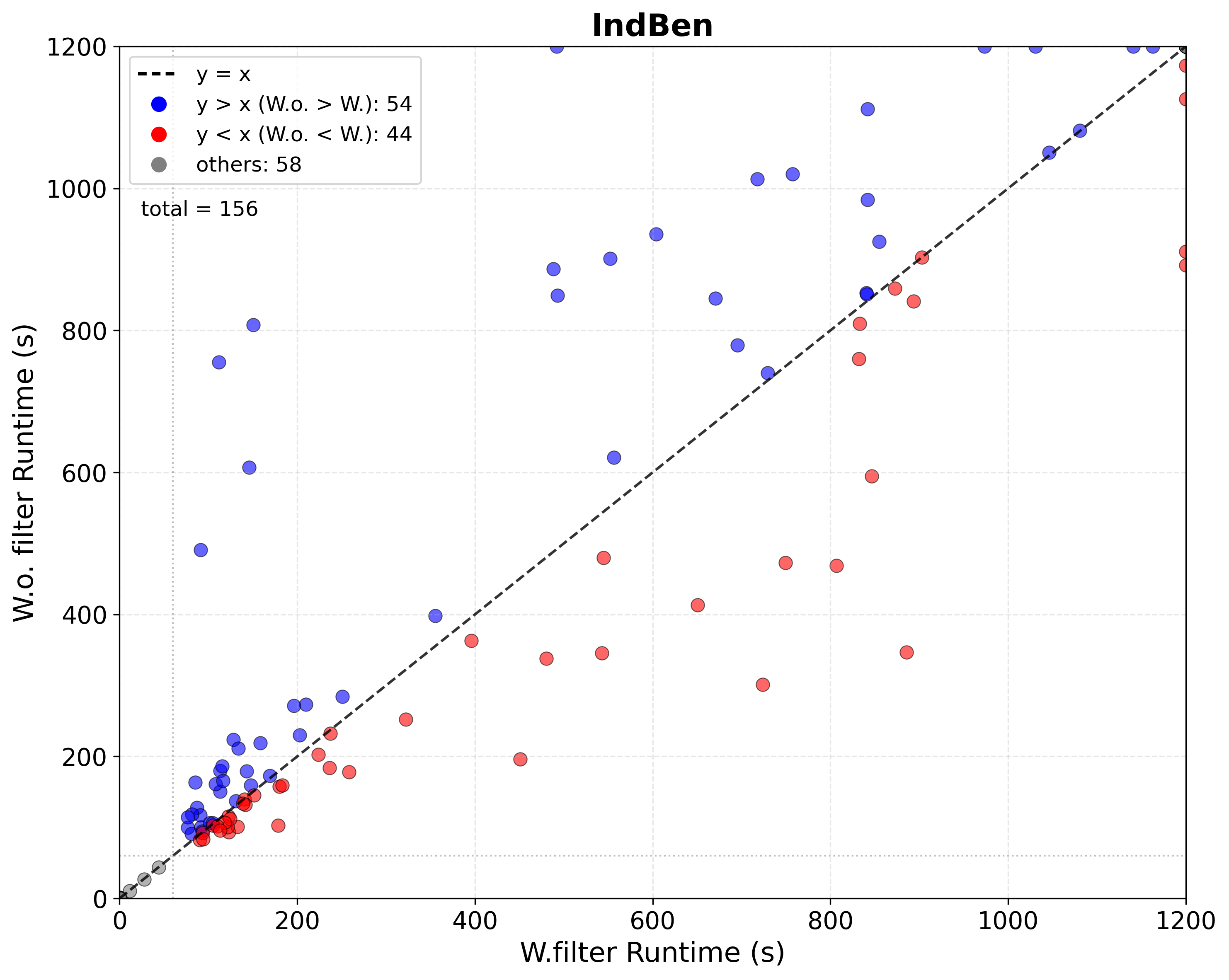}
\caption{IndBen benchmark}
\label{fig:time-comparison-indben}
\end{minipage}
\vspace{-4mm}
\end{figure}

To provide a detailed analysis of the filter's impact on solving time, we present scatter plots comparing the solving time of \tool with and without the filter across four benchmarks in Figures~\ref{fig:time-comparison-standarddt}--\ref{fig:time-comparison-indben}.

For each task we compute the average solving time over its three runs separately for the ``with the filter'' and ``without the filter'' settings. Unlike the main text, where only solved instances are included in the average, here we report the average time per task over its three runs (so that both settings can be compared on the same set of tasks). When a run hits the timeout, we record that run's time as 1200s.

The diagonal line is the equal-time line (solving time with the filter = solving time without the filter). The x-axis is the average time with the filter and the y-axis the average time without the filter. Points \emph{above} the diagonal (where time without the filter $>$ time with the filter) are colored \textbf{blue}, indicating instances where the filter reduces solving time. Points \emph{below} the diagonal (where time without the filter $<$ time with the filter) are colored \textbf{red}. In our setup, we first attempt to solve each instance by the SMT solver with a 60s time limit; only if it is not solved within 60s do we query the LLM. For a clearer view, tasks whose average time is below 60s (solved in the initial SMT phase) or equal to 1200s (timeout in every run) are shown in \textbf{gray}.

We can observe that across all four benchmarks, blue points outnumber red points, indicating that for most tasks the filter reduces or does not increase solving time and thus confirming the effectiveness of our filtering design.

\section{The Templates of two Prompt Strategies}

The following two listings show the prompt templates we use to guide LLMs 
for generating auxiliary lemmas in Section~\ref{ssec:prompt}.
\smallskip
\smallskip

\begin{adjustbox}{width=0.95\linewidth,center}
\begin{minipage}{\linewidth}
\begin{lstlisting}[language=SMT, style=Prompts,
  label=lst:designed-prompts1, belowcaptionskip=-25pt,caption={The template of Prompt Strategy 1}]
[Task Description]

You are an expert in constraint solving, inductive reasoning, and functional program verification. 
You are good at extracting information from SMTLIB2 files, reasoning about them, and generating necessary conjectures as auxiliary lemmas to help SMT solvers complete automatic proofs.
- Input format:
The input is an SMTLIB2 file. It contains: datatypes, function definitions, and the proof goal, each marked with ";" comments.
  
[Chain-of-Thought]

Please generate auxiliary lemmas using inductive equational reasoning step by step:
1) Identify the proof goal and list relevant axioms.
2) Assume the SMT solver already knows the induction scheme; you need to generate auxiliary lemmas to help the inductive reasoning.
3) Inductive proof setup:
   - Determine whether the base case requires auxiliary lemmas.
   - Derive the inductive case using equational reasoning.
4) Equational reasoning:
   - Transform the left-hand side of the property step by step, annotating axiom/hypothesis usage.
   - When a step cannot be derived, generate it as a conjecture, annotated as "unknown conjecture", which will be checked by SMT solvers to ensure it is an auxiliary lemma.

[Output Format]

- Please output all the "unknown conjectures" you discover through equational reasoning, but do not generate too many conjectures (at most 3 is recommended).
- Do not generate conjectures that are identical to the original property.
- Output each "unknown conjecture" in SMTLIB2 format on a single line.

[Input file]

{ Input SMTLIB2 file }
\end{lstlisting}
\end{minipage}
\end{adjustbox}

\begin{adjustbox}{width=0.95\linewidth,center}
\begin{minipage}{\linewidth}
\begin{lstlisting}[language=SMT, style=Prompts,
  label=lst:designed-prompts2, belowcaptionskip=-25pt,caption={The template of Prompt Strategy 2}]
[Task Description]

You are an expert in constraint solving, inductive reasoning, and functional program verification. 
You are good at extracting information from SMTLIB2 files, reasoning about them, and generating necessary conjectures as auxiliary lemmas to help SMT solvers complete automatic proofs.
- Input format:
The input is an SMTLIB2 file. It contains: datatypes, function definitions, and the proof goal, each marked with ";" comments.
  
[Chain-of-Thought]

Please generate auxiliary lemmas using the following ideas:
1) Generate basic axioms to help the SMT solver simplify the proof goal.
2) Strengthen the proof goal, e.g., to prove the proof goal P under axioms A, find a stronger conclusion Q, such that A => Q and Q => P hold, and both are easier to prove than A => P.
3) Based on term rewriting to simplify the proof goal:
   - Try to find the pattern term of the proof goal, and rewrite it to a simpler form.
   - If no common term exists, try to rewrite terms in the proof goal using axioms to have a common term.
   - If you think the auxiliary lemma derived from the term rewriting is not sufficient, generate a new auxiliary lemma that bridges it to the proof goal. 
4) Identify the conjectures from the above chain of thought, annotated as "unknown conjectures", which will be checked by SMT solvers to ensure they are auxiliary lemmas.
  
[Output Format]

- Please output all the "unknown conjectures" you discovered through reasoning, but do not generate too many conjectures (at most 3 is recommended).
- Do not generate conjectures that are identical to the proof goal.
- Output each "unknown conjecture" in SMTLIB2 format on a single line.

[Input file]

{ Input SMTLIB2 file }
\end{lstlisting}
\end{minipage}
\end{adjustbox}

}{}
\end{document}